\documentclass{ws-ijmpa}

\usepackage{epsfig}
\usepackage{amsmath}

\begin{document}

\markboth{M. B. Gay Ducati, Werner K. Sauter}
{Gluon propagator in diffractive scattering}

\catchline{}{}{}{}{}
%
%%%%%%%%%%%%%%%%%%%%%%%%%%%%%%%%%%%%%%%%%%%%%%%%%%%%%%%%%%%%%%%%%%%%

\title{Gluon propagator in diffractive scattering}

\author{M. B. Gay Ducati}
\address{Grupo de Fenomenologia de Part\'{\i}culas de Altas Energias,\\
Instituto de F\'{\i}sica, Universidade Federal do Rio Grande do Sul,\\
Caixa Postal 15051, CEP 91501-970, Porto Alegre, RS, Brasil\\
gay@if.ufrgs.br}

\author{W. K. Sauter}
\address{Grupo de Fenomenologia de Part\'{\i}culas de Altas Energias,\\
Instituto de F\'{\i}sica, Universidade Federal do Rio Grande do Sul,\\
Caixa Postal 15051, CEP 91501-970, Porto Alegre, RS, Brasil\\
sauter@if.ufrgs.br}

\maketitle

\begin{history}
\received{Day Month Year}
\revised{Day Month Year}
\end{history}

\begin{abstract}
In this work, we perform a comparison of the employ of distinct gluon propagators with the experimental data in diffractive processes, $pp$ elastic scattering and light meson photo-production. The gluon propagators are calculated through non-perturbative methods, being justified their use in this class of events, due to the smallness of the momentum transfer. Our results are not able to select the best choice for the modified gluon propagator among the analyzed ones, showing that the application of this procedure in this class of high energy processes, although giving a reasonable fit to the experimental data, should be taken with same caution.

 show  of the results for gluon propagator  can perform with care, .

\end{abstract}

\ccode{PACS numbers: 12.40.Nn, 13.85.Dz, 13.85.Lg, 14.40.Cs, 14.70.Dj}

%%%%%%%%%%%%%%%%%%%%%%%%%%%%%%%%%%%%%%%%%%%%%%%%%%%%%%%%%%%%%%%%%%%%%%%%%%%%%%%%%%%%%%%%%%%%
\section{Introduction \label{sec:intr}}
%%%%%%%%%%%%%%%%%%%%%%%%%%%%%%%%%%%%%%%%%%%%%%%%%%%%%%%%%%%%%%%%%%%%%%%%%%%%%%%%%%%%%%%%%%%%

The challenging description of the diffractive processes in high energy collisions in the framework of Quantum Chromodynamics has been requiring the attention over the last three decades\cite{fr97,bp02,ddln02}.  The main object of study is the Pomeron, which was originally proposed in Regge theory to explain the hadron collisions in the high energy limit. With the advent of QCD, the Pomeron was described by the exchange of a pair of gluons, in a first approximation or a ladder of gluons, the BFKL Pomeron\cite{fr97}.

However, in the kinematic region of small momentum transfer (called infrared region), we have divergences, due to the Landau poles in the gluon propagator. To remedy this problem, Landshoff and Nachtmann\cite{ln87} proposed a model for the Pomeron, in which the perturbative gluon propagator is replaced by a modified one in the infrared region by the inclusion of non-perturbative effects. Following the years, this model (or at least the main idea) was employed in several applications in diffractive scattering, as, for example, $pp$ elastic scattering, light meson photo-production.

As we review in the next section, there are several solutions for the gluon propagator in the literature, each one with a particular choice of parameters (specially the massive ones). In this work, we choose to employ mainly the solutions which have a dynamic gluon mass, which can have, or not, a finite mass for the gluon when the momentum is zero, but matches the perturbative behavior in large momentum transfer. There are also a direct connection of the gluon propagator\cite{amn02,amn04} and the frozen of the running coupling constant, which means a finite value of $\alpha_s$ for null momentum scale.

The {\em leitmotiv} of this work is to compare the different forms of the gluon propagator in an independent framework: the diffractive processes. There is a possibility to perform just a fit to the data, adjusting the massive parameters, however the idea here is to compare the original values found for the parameters, to make a true comparison among the propagators. This approach is supported by the previous attempts in this direction (see, for example, Refs.~\refcite{amn02,amn04,hkn93,gdws01} and references therein).

In the section \ref{sec:gpas}, we briefly review the methods to obtain the modified gluon propagators and presents the solutions with a dynamic gluon mass. The models to describe the diffractive scattering collisions are presented in the section \ref{ref:dsp} and their results are shown and discussed in the section \ref{sec:res}. We present our conclusions in section \ref{sec:rscl}.

%%%%%%%%%%%%%%%%%%%%%%%%%%%%%%%%%%%%%%%%%%%%%%%%%%%%%%%%%%%%%%%%%%%%%%%%%%%%%%%%%%%%%%%%%%%%
\section{The Gluon propagator and the running coupling constant \label{sec:gpas}}
%%%%%%%%%%%%%%%%%%%%%%%%%%%%%%%%%%%%%%%%%%%%%%%%%%%%%%%%%%%%%%%%%%%%%%%%%%%%%%%%%%%%%%%%%%%%

There are several methods to obtain the gluon propagator by using non-perturbative methods. The more popular are the Dyson-Schwinger equations and the lattice field theory.

The DSE's are an infinite system of non-linear coupled integral equations which relate the different Green functions of a quantum field theory. In the case of QCD, due to the complexity of the system to be solved, there are several analytical approximations in the literature to obtain a solution. A discussion among the different methods and approximations employed can be found in Refs.~\refcite{rw94,as01}.

The lattice field theory, in opposition to the DSE, is a powerful numerical method, where the main idea is that the divergences of the theory are regularized by the discretization of the space-time. Otherwise, the lattice approximation is not free of problems: finite size of the lattice, spacing between the sites of the lattice and fermion simulation. 

Due to the approximations employed in the resolution of the problems pointed above, there are many different solutions for both methods in the literature. We focus here on the solutions which have the particular property of a dynamical gluon mass: a gluon propagator with this feature was used in successful descriptions of several processes\cite{amn02,amn04,gds03}. 
The first solution for the gluon propagator which will be used in this work is the Cornwall's solution\cite{jmc82}, from DSE's in the axial gauge with the use of a technique of re-summation of Feynman diagrams,
\begin{equation}
D^{-1}_{C}({\mathbf q}^2) = \left[ {\mathbf q}^2 + m^2_C({\mathbf q}^2) \right] b g^2 \ln \left( \frac{{\mathbf q}^2 + 4 m^2_C({\mathbf q}^2)}{\Lambda^2_{\rm QCD}} \right), \label{e:dgco}
\end{equation}
with 
\begin{equation}
m^2_C({\mathbf q}^2) = m^2_g \left[ \ln\left(\frac{{\mathbf q}^2 + 4m^2_g}{\Lambda^2_{\rm QCD}}\right)\left/ \ln\left( \frac{4m^2_g}{\Lambda^2_{\rm QCD}} \right)\right. \right]^{-12/11},
\end{equation}
where $b=33/(48\pi^2)$ is the leading order coefficient of the $\beta$ function and $g=4\pi\alpha_s(0)$, with $\alpha_s$ the running coupling constant. The value for $m_g$ lies in the interval 500 $\pm$ 200 MeV for $\Lambda_{\rm QCD}$ = 300 MeV, given by the method of solution\cite{jmc82}. Another solution is obtained by H\"abel {\it et al.}\cite{hab90,hab90a} using a different type of approximation, which gives the following propagator,
\begin{equation}
D_{H}({\mathbf q}^2) = \frac{{\mathbf q}^2} {{\mathbf q}^2+b^4}, \label{e:dgh}
\end{equation}
where $b$ is a parameter (for its value, see below). Gorbar and Natale\cite{gn00} use the propagator
\begin{equation}
\left[D_{GN}({\mathbf q}^2)\right]^{-1} = {\mathbf q}^2 + \mu^2_g \Theta(\xi^\prime\mu^2_g-{\mathbf q}^2) + \frac{\mu^4_g}{{\mathbf q}^2}\Theta({\mathbf q}^2-\xi^\prime\mu^2_g) \label{e:dggn}
\end{equation}
to calculate the vacuum QCD energy through effective potentials, where $\Theta$ is the step function, $\mu_g =$ 0.61149 GeV is related with the gluon condensate and $\xi^\prime=$ 0.9666797 is a calculated parameter. A recent result for the gluon propagator, obtained from the DSE's in the Mandelstam approximation, is due to Aguilar and Natale\cite{an04a}, is given by the expression
\begin{equation}
D_{AN}({\mathbf q}^2) = \frac{1}{{\mathbf q}^2+\mathcal{M}^2(\mathbf{q}^2)},\quad\mathcal{M}^2(\mathbf{q}^2)=\frac{m_0^4}{\mathbf{q}^2+m_0^2}, \label{e:dgan}
\end{equation}
where $m_0^2 = 0.99\;{\rm GeV}^2$ for the QCD scale $\Lambda_{\rm QCD} =$ 335 MeV. 

Another recent result is the solution of DSE's in the Landau gauge calculated by Alkofer and collaborators\cite{as01,sha97,fa03,adfm04}, where the gluon propagator is given by \cite{adfm04}
\begin{equation}
D_{AL}({\mathbf q}^2) = \frac{\omega}{{\mathbf q}^2} \left[\frac{{\mathbf q}^2}{\Lambda^2_{\rm QCD}+{\mathbf q}^2}\right]^{2\kappa}\left(\alpha_s^{\rm (AL)}({\mathbf q}^2)\right)^{-\gamma},
\label{e:dgal}
\end{equation} 
where $\omega=2.5$, $\Lambda_{\rm QCD} = 510$ MeV, $\kappa \approx 0.595$, $\gamma = -13/22$ and $\alpha_s^{(\rm AL)}({\mathbf q}^2)$ is the running coupling constant,
\begin{equation}
\alpha_s^{(\rm AL)}({\mathbf q}^2) = \frac{1}{1+({\mathbf q}^2/\Lambda^2_{\rm QCD})} \left[ \alpha_s(0) + \frac{4\pi}{\beta_0} \frac{{\mathbf q}^2}{\Lambda^2_{\rm QCD}} \left( \frac{1}{\ln({\mathbf q}^2/\Lambda^2_{\rm QCD})} + \frac{1}{1-({\mathbf q}^2/\Lambda^2_{\rm QCD})} \right)  \right],
\label{e:arcc}
\end{equation}
where $\alpha_s(0) \approx 2.972$ and $\beta_0 = 11$.

%############################################################
\begin{figure}[t]
\begin{center}
\begin{tabular}{cc}
\scalebox{0.35}{\includegraphics*[25,35][510,520]{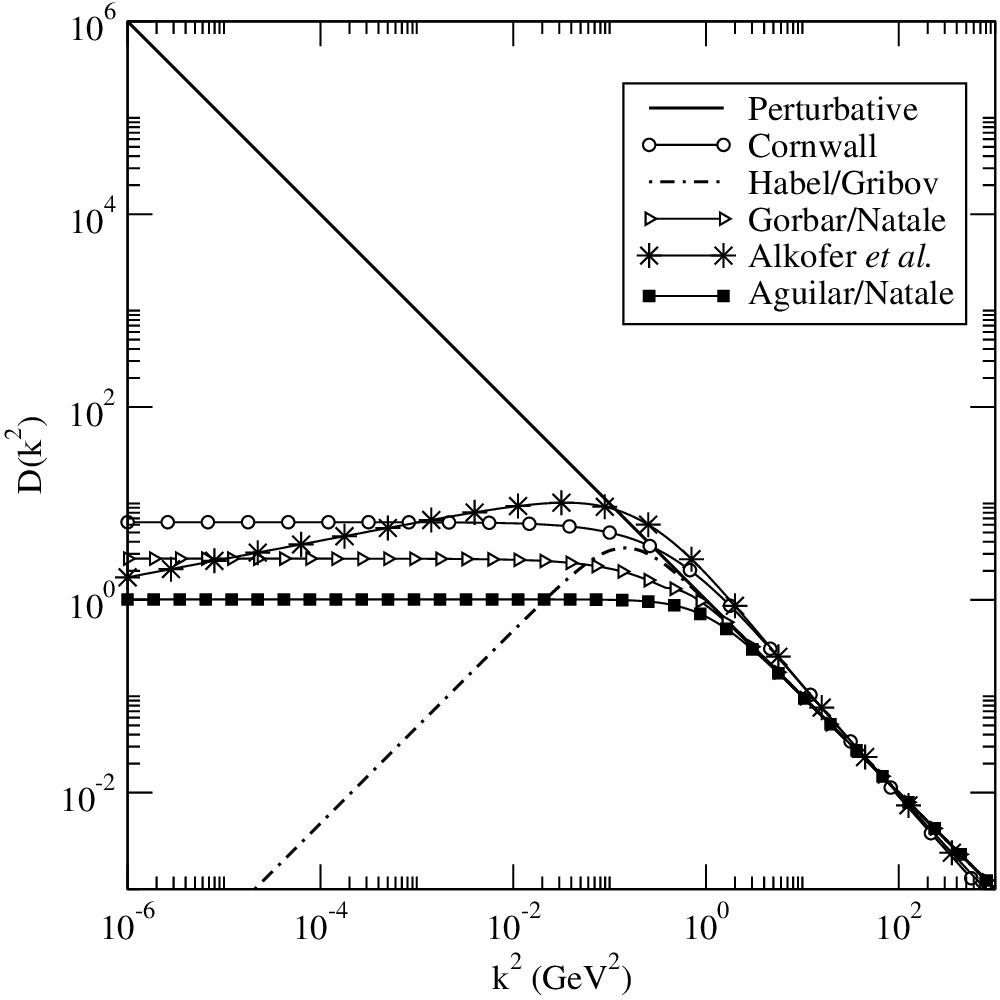}} &
\scalebox{0.35}{\includegraphics*[25,35][510,520]{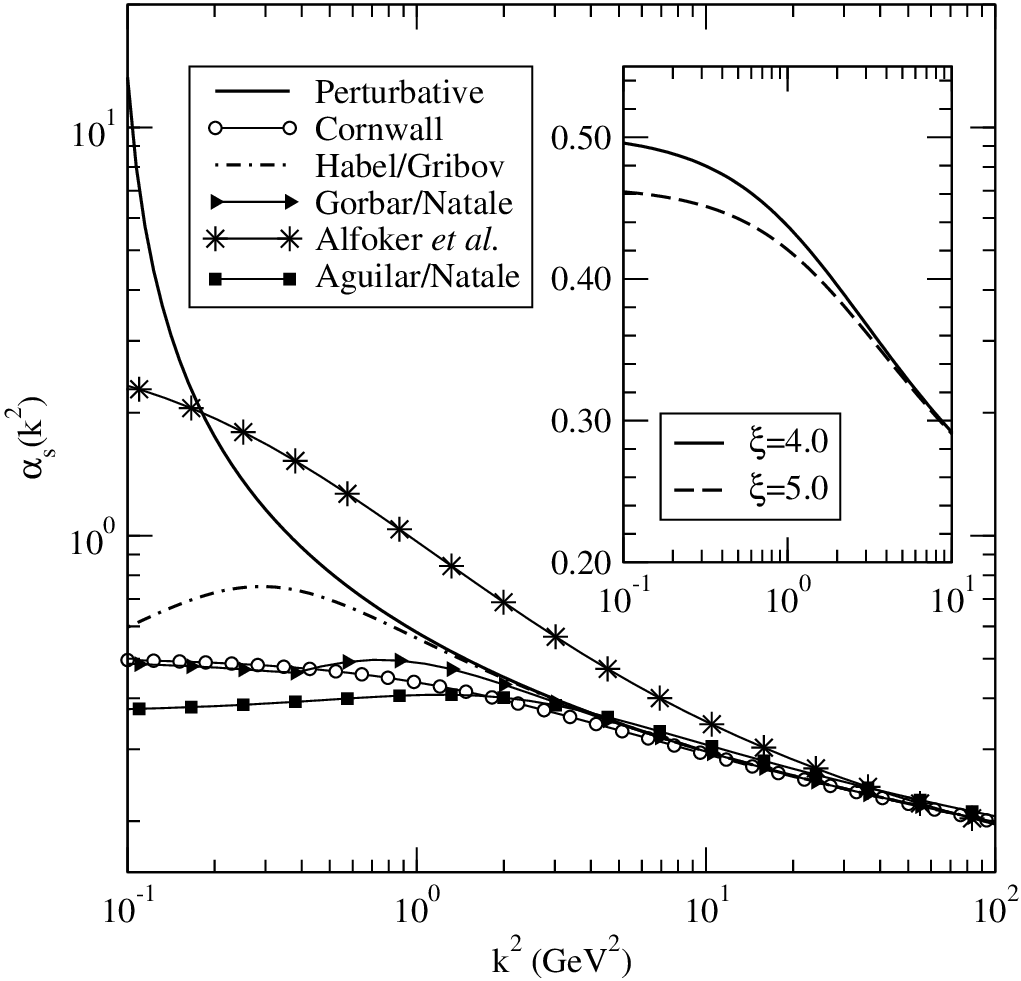}} \\
(a) & (b)
\end{tabular}
\end{center}
\caption{(a) Distinct modified gluon propagators compared with the perturbative one. The propagators are indicated in the plot legend. (b) Comparison between different results for the running coupling constant. In the detail, two results of the Cornwall frozen coupling with two different $\xi$'s. The parameters of the curves are given in the text.}  \label{f:dg}
\end{figure}
%############################################################

The momentum dependences of the gluon propagators are displayed in the figure \ref{f:dg}, in comparison with the perturbative one. The most remarkable feature is the infrared behavior ($\mathbf{k}\rightarrow 0$), where exists a finite value in the case of the Cornwall propagator, a zero value, as in the H\"abel/Gribov propagator, and the divergent perturbative propagator. Another remarkable fact is the same ultraviolet behavior ($\mathbf{k}\rightarrow \infty$) of the modified propagators and the perturbative one.

Nevertheless, there is a narrow relation between the gluon propagator and the running coupling, widely discussed in Refs.~\refcite{amn02,amn04}. The phenomenological applications of a frozen coupling constant include heavy quarkonia decays\cite{mn00}, meson form factors\cite{amn02,gds03} and other observables\cite{amn04}. In some of the above results (specially in Ref.~\refcite{gds03}), the data available are described when the modified gluon propagator and the frozen coupling constant are used at the same time.

The analytical expression for the running coupling is obtained following the Cornwall's solution\cite{jmc82,cp91}, namely,
\begin{equation}
\alpha_s(k^2)=4\pi\left/ \beta_0\ln\left(\frac{k^2+\xi m_{\mathcal P}^2(k^2)}{\Lambda^2_{\rm QCD}}\right) \right. ,
\label{e:frcc}
\end{equation}
where the explicit momentum dependence is the same as the one loop perturbative calculation, but with an extra term, $\xi m_{\mathcal P}^2(k^2)$, resulting in the frozen infrared behavior where $\xi$ is a parameter that varies with the propagator and measures the frozenness of the coupling. In the original work\cite{jmc82,cp91}, the value of this parameter is obtained as $\xi\approx 4$ or larger.  The massive term, $m_{\mathcal P}(k^2)$, also depends on the gluon propagator employed, being the dynamical mass term of the propagator. For example, in the case of the Aguilar and Natale propagator, eq. (\ref{e:dgan}), this term is $m_{\mathcal P}(k^2)=\mathcal{M}^2(k^2)$. The exception is the Alkofer case, with a proper expression for the running coupling constant, eq. (\ref{e:arcc}).

In the fig.~\ref{f:dg}, we present the different $\alpha_s$'s (given by eq.(\ref{e:frcc})), where the freezing of the running coupling constant, the matching of the frozen running coupling, and the behavior of the perturbative result in high momentum are very clear. The last result comes from the behavior of the massive term ($m_{\mathcal P}(k^2)$) which goes to zero when the momentum is large in all the propagators employed in this work. The values of the propagator parameters are from the original works or assumed in the region of validity of the models, as in case of the Cornwall propagator. The figure also shows the $\xi$ parameter dependence of the Cornwall solution, eq. (\ref{e:dgco}). As seeing in fig.~\ref{f:dg}, there is a small difference for distinct $\xi$ values.

%%%%%%%%%%%%%%%%%%%%%%%%%%%%%%%%%%%%%%%%%%%%%%%%%%%%%%%%%%%%%%%%%%%%%%%%%%%%%%%%%%%%%%%%%%%%
\section{Diffractive scattering processes} \label{ref:dsp}
%%%%%%%%%%%%%%%%%%%%%%%%%%%%%%%%%%%%%%%%%%%%%%%%%%%%%%%%%%%%%%%%%%%%%%%%%%%%%%%%%%%%%%%%%%%%

The framework in which the different results for the gluon propagator are compared is the diffractive scattering. This class of QCD processes is characterized by the exchange of the Pomeron and the consequent rapidity gaps in the data\cite{bp02}. In the regime of small momentum transfer, it is a good laboratory to test the modified gluon propagators, due to the range of validity of the approximation employed in the derivation of these objects. 

The most employed model in this case is the Landshoof-Nachtmann one\cite{ddln02}, where the Pomeron is modeled by the exchange of two non-perturbative gluons (with modified gluon propagators). In the original version of the LN model, the parameters are fixed relating the Pomeron strength radius and the Pomeron effective mass. A disadvantage of the LN model is the energy independent cross sections, due to the two gluon approximation.

We will analyze two processes in which the small momentum contributions are relevant and are mediated by the Pomeron exchange: the $pp$ scattering and the electro-production of light mesons; althought already present in the literature, not with the emphasis assumed in this present work.

The model employed to describe the proton-proton elastic scattering in this work was proposed by Cudell and Ross\cite{cr91}, based on the LN two-gluon model for the Pomeron. A feature of this model is the strong dependence on the wave functions of the involved hadrons, giving a possible source of uncertainty. The LN model is used to avoid the infrared divergences employing a modified gluon propagator, with the forms displayed above.

The scattering amplitude can be written as\cite{cr91} 
\begin{equation}
\mathcal{A}^{pp}(s,t) = 8is\alpha_s^2\left({\mathcal T}_1-{\mathcal T}_2\right), \label{eq:app2}
\end{equation}
where
\begin{subequations}
\begin{eqnarray}
\mathcal{T}_1 &=& \int\!d^2\mathbf{k}\; \mathcal{D}\left(\frac{\mathbf{q}}{2}+\mathbf{k}\right) \mathcal{D}\left(\frac{\mathbf{q}}{2}-\mathbf{k}\right) G_p^2(q,0), \label{e:sct1} \\
\mathcal{T}_2 &=& \int\!d^2\mathbf{k}\; \mathcal{D}\left(\frac{\mathbf{q}}{2}+\mathbf{k}\right) \mathcal{D}\left(\frac{\mathbf{q}}{2}-\mathbf{k}\right) G_p\left(q,k-\frac{q}{2}\right) \left[ 2G_p(q,0) - G_p\left(q,k-\frac{q}{2} \right)\right], \label{e:sct2}
\end{eqnarray}
\end{subequations}
where $s,t=-q^2$ are the Mandelstam variables and $G_p(q,k)$ is a convolution of the proton wave functions,
\[ G_p(q,k) = \int\!d^2p\,d\alpha\;\psi^\ast(\alpha,p)\;\psi(\alpha,p-k-\alpha q), \]
and can be related with the Dirac form factor as follows,
\begin{subequations}
\begin{eqnarray}
G_p(q,0) &=&  F_1(q^2) \label{e:ff1} \\
G_p\left(q,k-\frac{q}{2}\right) &=& F_1\left(q^2+9\left|k^2-\frac{q^2}{4}\right|\right)
\end{eqnarray}
\end{subequations}
where 
\begin{equation}
 F_1(t) = \frac{4m_p^2-2.79t}{(4m_p^2-t)\left(1-t/0.71\right)^2}, \label{eq:dff}
\end{equation}
where $m_p$ is the proton mass. 

The terms in eqs.~(\ref{e:sct1},\ref{e:sct2}) have a simple interpretation: the former comes from Feynman diagrams where the gluons are connected to the same quark in the proton. The last one comes from diagrams where the gluons are connected to different quarks in the proton\cite{cr91}.

The total cross section is given by the optical theorem, $\sigma_{\rm tot}^0 = \mathcal{A}_2^{pp}(s,0)/is$, as well as the differential cross section $d\sigma^0/dt = |\mathcal{A}_2^{pp}(s,t)|^2/16\pi s^2$. For low momentum transfer, the elastic differential cross section is fitted by an exponential expression, $d\sigma/dt = A\,e^{Bt} = \sigma_{\rm tot}^2\,e^{Bt}/(16\pi)$, where $B$ is the logarithmic slope, given by $B=d/dt[\ln(d\sigma/dt)]|_{t=0}$. 

With the scattering amplitude above, a finite result for the total cross section is obtained due to the cancellation of the infrared divergences\cite{hkn93,cr91}. Otherwise, the differential cross section has an unphysical result ($B(t=0)\rightarrow\infty$) if the perturbative gluon propagator is employed. When the Landshoff-Nachtmann model is used, the contribution from the term ${\mathcal T}_2$ is negligible and the remaining infrared divergences are regularized by the modified gluon propagators.

However, the two gluon description of the scattering amplitude is a crude approximation, which does not yield any energy dependence in the above cross sections. The reason is the absence of the gluon ladders in the $s$ channel, as in the BFKL framework, which gives an energy dependence in the cross sections. To restore this dependence, we assume that the amplitude above only describes the energy independent part of the cross section and an extra term, introduced {\it ad hoc} gives the Regge energy behavior, namely,
\[ \mathcal{A}^{pp}(s,t) \rightarrow \left(\frac{s}{s_0}\right)^{\alpha_{P}(t)-1}\mathcal{A}^{pp}(s,t), \]
where $\alpha_{P}(t)=\alpha(0) + \alpha^\prime(t)$ is the (soft) Pomeron trajectory (with\cite{dl92} $\alpha(0)=1.08$ and $\alpha^\prime(t) \simeq 0.25$ GeV$^{-2}$) and $s_0$ is an energy scale. The extra Regge term gives the following cross sections,
\begin{subequations}
\begin{eqnarray}
\sigma_{\rm tot} &=& \left(\frac{s}{s_0}\right)^{\alpha_{P}(0)-1}\sigma_{\rm tot}^0, \label{e:tostf} \\
\frac{d\sigma}{dt} &=& \left(\frac{s}{s_0}\right)^{2\alpha_{P}(t)-2}\frac{d\sigma^0}{dt} \label{e:todsf}.
\end{eqnarray}
\end{subequations}

The above model was used by Refs.~\refcite{hkn93,cr91,hpr96} where the non-perturbative propagator comes from different methods, discussed in the next session. The experimental data are ISR results for elastic proton-proton scattering at\cite{isr84} $\sqrt{s}=$ 53 GeV.

In the previous model for the scattering amplitude, the running coupling constant was considered fixed. In Ref.~\refcite{gdhn93}, the value of the coupling was determined thought the effective Pomeron coupling with the hadrons, which depends on the gluon propagator and has well defined experimental value.

Motivated by the good results of our previous calculations with a gluon propagator and a frozen running coupling\cite{gds03}, we modify the former model of $pp$ scattering, considering now a frozen running coupling constant and a modified gluon propagator. We emphasize that this is a phenomenological modification, since a running coupling is a next leading order contribution and the model employed is a first order contribution to the Pomeron exchange. As usual in previous works, the scale of the running coupling constant is chosen as the incoming momentum into the quark-gluon vertex.

After the above considerations, the scattering amplitude in the modified model then reads as
\begin{eqnarray}
\mathcal{A}^{pp}_2(s,t) &=& 8is\left(\frac{s}{s_0}\right)^{\alpha_{P}(t)-1}\int\!d^2\mathbf{k}\; \alpha_s\left(\frac{\mathbf{q}}{2}+\mathbf{k}\right) \mathcal{D}\left(\frac{\mathbf{q}}{2}+\mathbf{k}\right) \alpha_s\left(\frac{\mathbf{q}}{2}-\mathbf{k}\right) \mathcal{D}\left(\frac{\mathbf{q}}{2}-\mathbf{k}\right) \times \nonumber \\
& &\left[ G_p(q,0) - G_p\left(q,k-\frac{q}{2}\right) \right]^2. \label{e:appm}
\end{eqnarray} 
where the terms from eqs. (\ref{e:sct1}) and (\ref{e:sct2}) were combined as only one term and the running coupling terms are included in the integrand. 

The production of massive vector mesons is a process in which the Pomeron assumes an important role\cite{ac99}, being a powerful laboratory for the study of the diffraction. The relevant process is
\[ \gamma^{(\ast)} + p \rightarrow V + p \]
where $V$ is the vector meson, $p$ is the proton and $\gamma$ is the photon, being real (photo-production) or virtual (lepto-production). The process can be described in the high energy regime in a factorized way in three steps: first, the photon splits into a pair of quark and anti-quark; second, the pair interacts with the proton through the exchange of a color singlet, the Pomeron (the meson has the same quantum numbers of the photon) and last, the anti-quark-quark pair recombines into a bound state, the vector meson. The scattering amplitude can be factorized in these three steps, as will be discussed below. Relevant kinematic variables are the photon virtuality, $Q^2$, the meson mass, $M^2_V$, the square of the exchanged momentum, $t$, and the center-of-mass energy squared of $\gamma p$, $W^2$.

In the following, we employ the model of Cudell and Royen\cite{rc97} in the description of the production of $\rho$ mesons in H1 and ZEUS, with the same modifications already used in the $pp$ elastic scattering, meaning the same set of modified gluon propagators and running coupling constant. 

\begin{figure}[t]
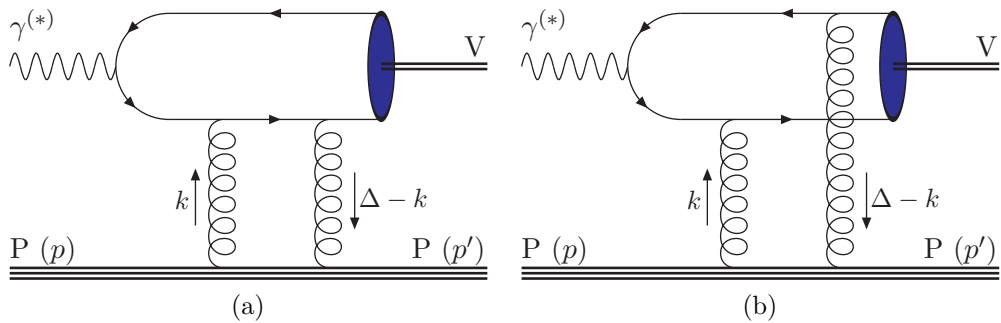
 
\begin{center} 
\begin{tabular}{cc}  
\epsfig{file=mesfig1.epsi} & 
\epsfig{file=mesfig2.epsi} \\
(a) & (b)
\end{tabular} \end{center}
\caption{Contributing Feynman diagrams to the meson vector production by two gluons exchange.}  
\label{f:dfey}
\end{figure}

A factorization scheme is employed, where three steps are considered, one of each with the following assumptions: (a) the interaction of the Pomeron with the proton, where the former interacts only with one valence quark of the proton, considering the Dirac elastic form factor for the proton; (b) the coupling of the Pomeron with the pair and the transition $qq\rightarrow V$, modeled assuming that the meson is in the lowest Fock state (two quarks) and without any Fermi momentum, allowing the use of a non-relativistic model for the $qqV$ vertex. However, Cudell's model\cite{jrc90} does not give any energy dependence for the cross sections and, for this reason, a Regge-like energy term is added {\it ad hoc}, due to the exchange of only two gluons. In this case, the (b) step above includes the possibility that the Pomeron (as two gluons) can be attached in different or in the same quark in the virtual pair, as shown in fig. \ref{f:dfey}. However, contributions from diagrams where the gluons of the Pomeron are attached with different quarks of the proton are suppressed in order to reproduce the quark-counting rule in the LN model\cite{ln87}. An improvement considering that the interaction of the two gluons with the proton comes from two Feynman diagrams, one in which the gluons are attached with the same quark in the proton and another in which the gluons are attached to different quarks was given in Ref.~\refcite{rc97}. The contributions of these diagrams eliminate the infrared divergences as pointed out by Ref.~\refcite{cn94}. The last contribution, the only one from attachment to different quarks, is modeled by a modification of the Dirac elastic form factor, as discussed below. 

%############################################################
%############################################################

In this paper, we consider the simple model of Cudell and Royen\cite{rc97}, with the same modifications used in the description of the $pp$ scattering, supported by several reasons. First, its simple form provides an easy and direct modification of the gluon propagator and running coupling constant. Thus, a modified gluon propagator and a frozen running coupling constant as well as a Regge term to restore the energy dependence are employed, supported by a series of previous works\cite{jrc90,gdhn93,hkn93,amn04}. This formulation has the goal to test the modified gluon propagators in association with a frozen running coupling constant, not present in any previous model to analyze this process. 

A discussion about the terms that contribute to the unmodified scattering amplitude can be founded in the original work\cite{rc97}. The amplitude for the possible allowed helicities (transverse to transverse and longitudinal to longitudinal, the other cases being suppressed by $w^{-2}$) is given by
\begin{eqnarray}
A_{TT} &=& iw^2\frac{64}{\sqrt{6}} \frac{g_{\rm elm}^{V} m_{V} \sqrt{m_{V}f_{V}}}{m^2_{V}+Q^2-t}  \\ \nonumber
& & \times \int \frac{\alpha^2_s\,d^2{\mathbf k}}{{\mathbf k}^2 ({\mathbf k}-{\mathbf \Delta})^2} \frac{[F_1(t)-{\mathcal E}_2(k,k-\Delta)][{\mathbf k}\cdot({\mathbf k}-{\mathbf \Delta})]}{t-m^2_V-Q^2+4{\mathbf k}\cdot({\mathbf k}-{\mathbf \Delta})}, \\
\label{e:ampl1}
A_{LL} &=& \frac{\sqrt{Q^2}}{m_{V}} A_{TT},
\label{e:ampl2}
\end{eqnarray}
where $w^2$ is the center-of-mass energy squared of the quark-photon system, $g_{\rm elm}^{V}=\xi\sqrt{4\pi\alpha_{\rm elm}}$ is the electromagnetic coupling of different vector mesons: $\xi=1/\sqrt{2}$ for the $\rho$ meson; $f_V=0.025 m_V$ is the vector meson decay constant squared, with $m_V$ the meson mass; $F_1(t)$ is the elastic Dirac form factor,
\[ F_1(t) = \frac{4m^2_p - 2.79t}{(4m_p^2-t)(1-t/0.71)^2}\ , \]
with $m_p$ the proton mass and ${\mathcal E}_2(k,k-\Delta) = F_1({\mathbf k}^2 + ({\mathbf k}-{\mathbf \Delta})^2-{\mathbf k}\cdot({\mathbf k}-{\mathbf \Delta}))$. The last term comes from the diagrams that couples with different quarks inside the proton and, in the original work, avoids the infrared divergences.

The differential cross section is\cite{rc97}
\begin{equation}
\frac{d\sigma}{dt} = \frac{d\sigma_T}{dt} + \epsilon\frac{d\sigma_L}{dt} = \frac{R}{16\pi w^4}\left\{|A_{TT}|^2+\epsilon|A_{LL}|^2 \right\},
\label{e:dcs}
\end{equation}
where $\epsilon$ is the polarization of the photon beam ($\epsilon\approx 1$ at HERA) and $R$ is a factor, needed to restore the energy dependence, since two gluon exchange does not give such dependence in the cross section.

The amplitude, eq. (\ref{e:ampl1}), now reads
\begin{eqnarray}
A^\prime_{TT} &=& iw^2\frac{64}{\sqrt{6}} \frac{g_{\rm elm}^{V} m_{V} \sqrt{m_{V}f_{V}}}{m^2_{V}+Q^2-t} \left(\frac{s}{s_0}\right)^{\alpha_{P}(t)-1} \int d^2{\mathbf k}\,\alpha_s({\mathbf k})\alpha_s({\mathbf k}-{\mathbf \Delta}) \nonumber \\  
& & \times {\mathcal D}({\mathbf k}) {\mathcal D}({\mathbf k}-{\mathbf \Delta}) \frac{[F_1(t)-{\mathcal E}_2(k,k-\Delta)][{\mathbf k}\cdot({\mathbf k}-{\mathbf \Delta})]}{t-m^2_V-Q^2+4{\mathbf k}\cdot({\mathbf k}-{\mathbf \Delta})} \label{e:ampl1m} \\
A^\prime_{LL} &=& \frac{\sqrt{Q^2}}{m_{V}} A^\prime_{TT}. \label{e:ampl2m}
\end{eqnarray}

The expression (\ref{e:ampl1m}) can be rewritten in a more compact way, 
\[ A^\prime_{TT} = iw^2{\mathcal C}\zeta^{\alpha_{P}(t)-1} f\left({\mathbf \Delta}, Q^2\right), \] where $\zeta=s/s_0$ with $s_0=Q^2+m_V^2-t$, ${\mathcal C}=(64/\sqrt{6})g_{\rm elm}^{V} m_{V} \sqrt{m_{V}f_{V}}$  and $f\left({\mathbf \Delta}, Q^2\right)$ covers the remaining terms. The differential cross section comes from the optical theorem,
\begin{equation}
\frac{d\sigma}{dt} = \frac{1}{16\pi}{\mathcal C}^2\zeta^{2(\alpha_{P}(t)-1)} f^2\left({\mathbf \Delta}, Q^2\right)\left[ 1+\epsilon\frac{Q^2}{m^2_V}\right], \label{e:dsdtm}
\end{equation}
as well as, the total cross section,
\begin{equation} 
\sigma_{\rm tot} = \int\!dt\,\frac{d\sigma}{dt} = \frac{1}{16\pi}{\mathcal C}^2\left[ 1+\epsilon\frac{Q^2}{m^2_V}\right] \int\!dt\,\zeta^{2(\alpha_{P}(t)-1)} f^2\left({\mathbf \Delta}, Q^2\right). \label{e:stm}
\end{equation}

%%%%%%%%%%%%%%%%%%%%%%%%%%%%%%%%%%%%%%%%%%%%%%%%%%%%%%%%%%%%%%%%%%%%%%%%%%%%%%%%%%%%%%%%%%%%
\section{Results} \label{sec:res}
%%%%%%%%%%%%%%%%%%%%%%%%%%%%%%%%%%%%%%%%%%%%%%%%%%%%%%%%%%%%%%%%%%%%%%%%%%%%%%%%%%%%%%%%%%%%

%############################################################
\begin{figure}[t]
\begin{center}
\begin{tabular}{cc}
\scalebox{0.35}{\includegraphics*[70,35][535,535]{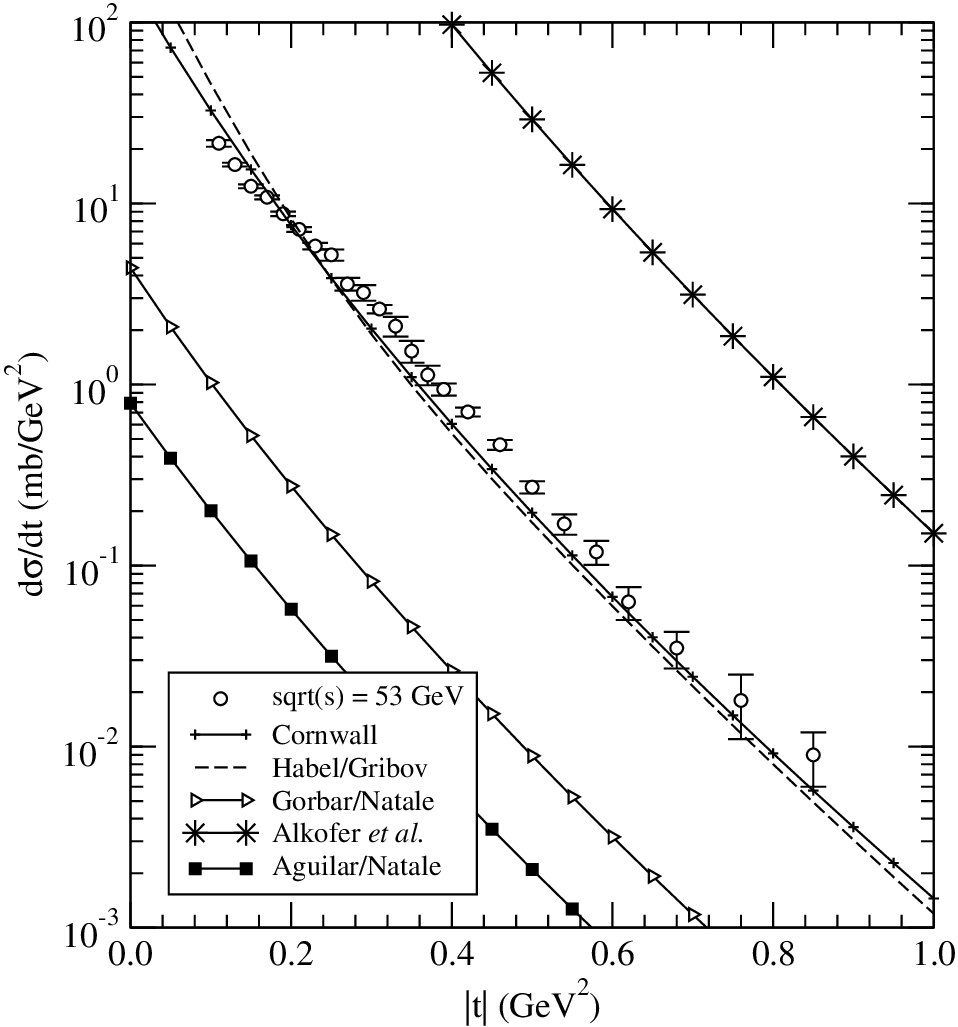}} & 
\scalebox{0.35}{\includegraphics*[70,35][535,535]{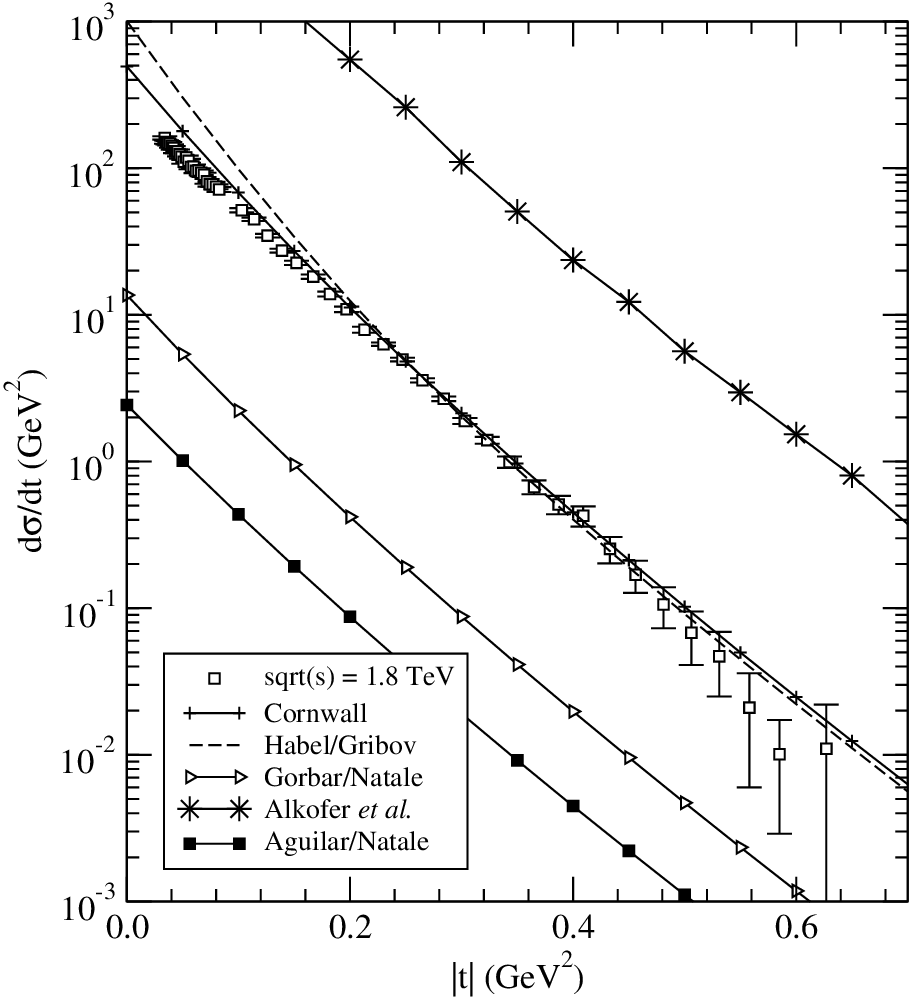}}
\end{tabular}
\end{center}
\caption{Results of the modified model, fitting to the $pp$ elastic scattering data with $\sqrt{s}=$ 53 GeV (right) and $\sqrt{s}=$ 1.8 TeV (left) with different gluon propagators and frozen coupling constants. The parameters of the fits are given in the text.}  
\label{f:dsdt}
\end{figure}
%############################################################

%############################################################
\begin{figure}[t]
\begin{center}
\begin{tabular}{cc}
\scalebox{0.35}{\includegraphics*[25,45][500,520]{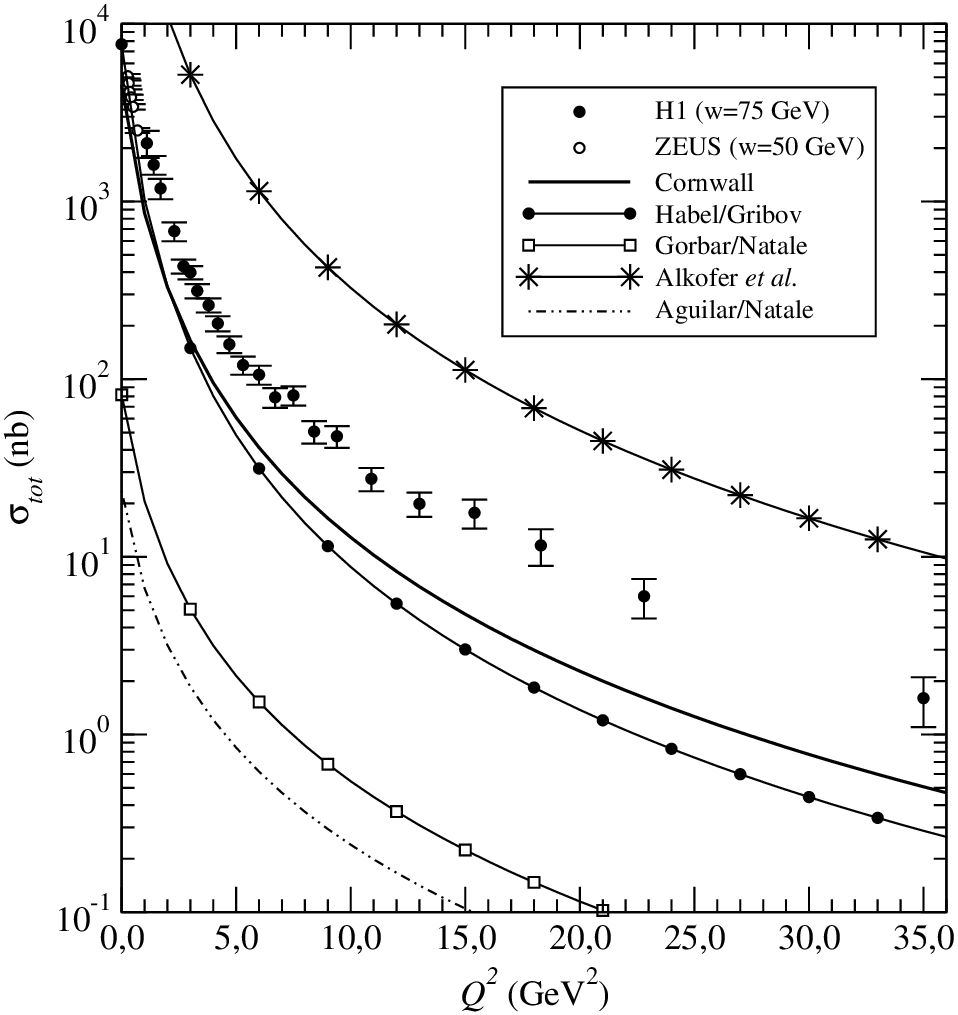}} &
\scalebox{0.35}{\includegraphics*[25,45][500,520]{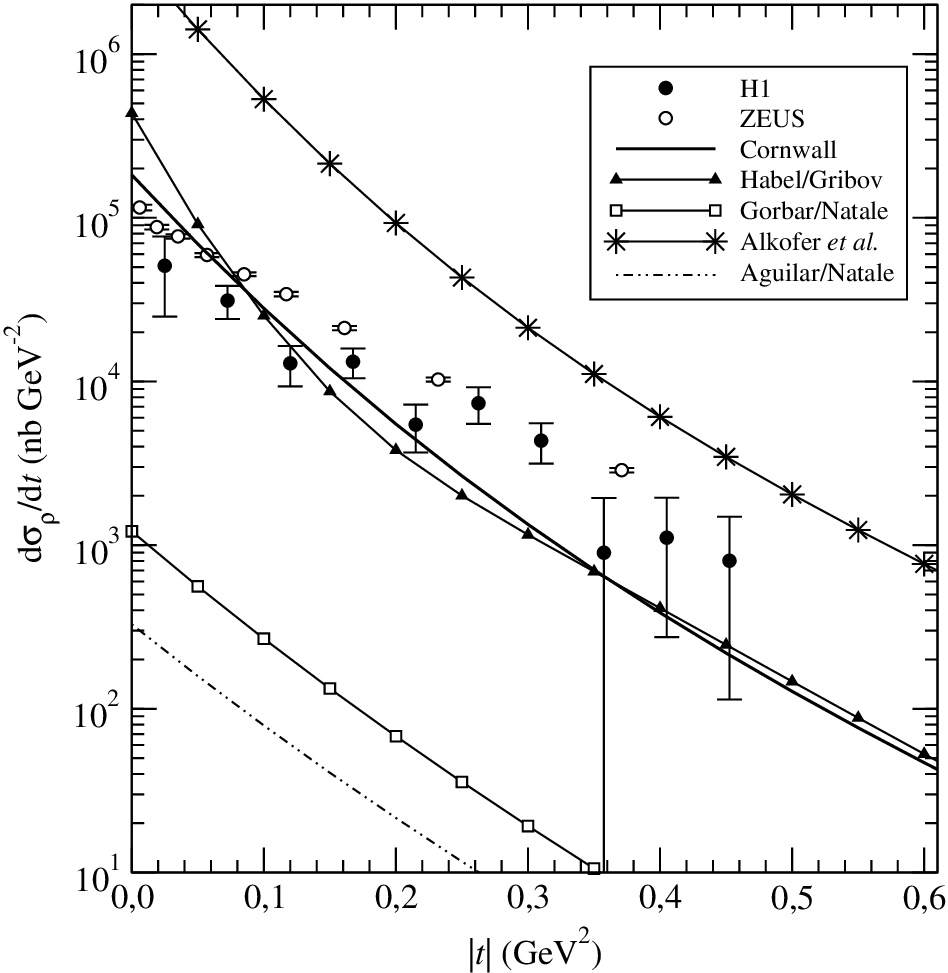}} \\
(a) & (b) 
\end{tabular}
\end{center}
\caption{(a) Total cross section obtained with the eq. (\ref{e:stm}) for $\rho$ meson. (b)$\rho$ photo-production differential cross section, compared with the H1 and ZEUS data.}
\label{f:stds}
\end{figure}
%############################################################

The differential cross section for $pp$ scattering is calculated using the eq. (\ref{e:appm}) into eq. (\ref{e:todsf}), varying the propagators as well as the running coupling constants, as pointed out above. The results of the calculation are displayed in the fig.~\ref{f:dsdt} for two sets of experimental data for distinct center of mass energies: $\sqrt{s}=$ 53 GeV from ISR\cite{isr84}, and $\sqrt{s}=$ 1.8 TeV from Tevatron\cite{e71090}.

Some comments over the results are addressed in the following. In the case of the $pp$ scattering, we see the significant variation in the cross section when different propagators are employed. The slope of the curves are in accordance with  data, due the {\it ad hoc} term in the amplitude. The energy independent part of the amplitude, modeled by the eq.~(\ref{eq:app2}) gives a contribution on the $t$ behavior, on the intercept ($d\sigma/dt|_{t=0}$) of the curve and the normalization of the result. If we allow the variation of the propagator parameters, is easy to fit the curves to the data, for both energies. The calculation is sensitive to the value of the parameters, giving larger modifications when the massive parameter is changed. In opposition, when the $\xi$ parameter is changed, the change in the result is significantly smaller than with the mass parameter. Thus, we will use the value of the $\xi$ fixed as 4. 

In the case of the Cornwall propagator, the propagator's parameter best value is $m_g= 0.53$ GeV, which lies in the range of validity of the approximation employed (see above), and for the H\"abel/Gribov propagator, the value for the parameter is $b=0.3$ GeV, close to the previous one, used in Ref.~\refcite{gdws01}. In all the other propagators, the parameters are the original ones. In two cases, the Gorbar/Natale one and Aguilar/Natale one, we obtain that the original choice of the parameters underestimate the results. When employing the Alkofer {\it et al.} propagator, eq. (\ref{e:dgal}), and the corresponding running coupling constant, eq. (\ref{e:arcc}), we obtain a result that overestimates the experimental data for both energies. Similar result was found by Refs.~\refcite{amn02,amn04} for another observables, supporting our results.

The results of the modified model for the total cross section, eq. (\ref{e:stm}), are displayed in figure \ref{f:stds}, for the gluon propagators and coupling constants discussed in the previous section and in comparison with the HERA (H1 and ZEUS) data\cite{h100,zs99}.

The center-of-mass energy is $w \approx$ 100 GeV, the result being less sensitive to this parameter. Due to the reasons exposed above, the energy dependent term uses the soft Pomeron trajectory, $\alpha_{P}(t)=\alpha(0) + \alpha^\prime(t)$, with $\alpha(0)=1.08$ and $\alpha^\prime(t) \simeq 0.25$ GeV$^{-2}$. As in the $pp$ elastic scattering case, the $\xi$ parameter of the running coupling constant is fixed in $\xi=4$, its change does not modify significantly the results.

Therefore, the gluon parameters are, for the cases with some freedom of choice: Cornwall, $m_g=0.4$ GeV and H\"abel/Gribov, $b= 0.2$ GeV. The final result is very sensitive to this parameter, giving very different results if altered by a small amount, allowing the fit of the data, if the parameters are free.

The results lie along in the same general features of the $pp$ scattering: if the original values of the propagator parameters are employed, the data are not fitted, with the exception of two propagators, which have some choice of the parameter. The overall behavior describes the data, although the normalization which depends on the choice of the propagator is large. However, if the propagator parameter is allowed to vary, we can improve the agreement of the fit, as in the $pp$ scattering, but in the case of the total cross section,  in the high-$Q^2$ region, the result stays above the data, showing that the model can describe better the kinematic region where the non-perturbative effects are relevant. The differential cross section of photo-production ($Q^2=0$) of the $\rho$ meson is displayed in the fig. \ref{f:stds}, where the differences between the results from different propagators are evident, but we have a reasonable description of the data. As in the proton-proton case, some of the propagators lie below the data and the Alkofer's choice lies above the cross section data.

%%%%%%%%%%%%%%%%%%%%%%%%%%%%%%%%%%%%%%%%%%%%%%%%%%%%%%%%%%%%%%%%%%%%%%%%%%%%%%%%%%%%%%%%%%%%
\section{Conclusions \label{sec:rscl}} 
%%%%%%%%%%%%%%%%%%%%%%%%%%%%%%%%%%%%%%%%%%%%%%%%%%%%%%%%%%%%%%%%%%%%%%%%%%%%%%%%%%%%%%%%%%%%

We present a comparison of different results for the gluon propagator and the related frozen running coupling constant, which functional forms are obtained from different non-perturbative methods. These results are tested in an appropriated framework, the high energy diffractive scattering, where the low momentum transfer effects are relevant, and therefore the application of the modified gluon propagator is relevant. Among the several processes, our choices are the elastic $pp$ scattering and light meson production. The same ideas were employed by the present authors, in different processes also with a small momentum transfer as, for example, the description of meson form factors\cite{amn02,gds03} and the process $\gamma\gamma\rightarrow J/\psi J/\psi$ also was described in a similar framework in Ref.~\refcite{gdws01}.

The original idea is not to perform a fit to the data, but confront the original solutions for gluon propagators through specific observables. The different values for the massive parameter obtained for the $pp$ elastic scattering process indicates some weakness in this approach. The same problems appear in the meson production, where the massive parameter is different of the $pp$ case. If the functional forms for the gluon propagator is correct, what is the origin of these discrepancies? An answer could be the model employed in the calculation, which have the same common origin, but then the results would be similar, which is not the case.

The results for the description of the observables are not quite satisfactory, showing that the approximation employed here, although still crude, signs that the use of the modified gluon propagator in diffractive events can be a reasonable description of the data, although not complete. The crudeness of the fit in some cases can indicate an incomplete picture for diffractive scattering, and also our choice of not performing a fit to the experimental data, adjusting the massive parameter, may be a very restrictive one. Moreover, the present results show that the assumptions of the modified model should be used with caution. The gluon propagators are obtained in a self consistent way, although the original choices for the parameters do not give good results. Otherwise, the model employed can be incomplete, omitting contributions from other Green functions, as quark propagators, for example. Thus, a possible natural extension is to employ a complete set of Green functions: the quark and gluon propagators and the quark-gluon vertex.

In the present processes, in which the choices have the feature of a dynamical mass (or a massive parameter), we obtain a quite good agreement with experimental data for distinct energies, for two choices for propagators, the Cornwall's and H\"abel/Gribov ones, which have some freedom to vary the parameters. The results obtained are along the same lines of the previous attempts to describe these processes, however considering distinct propagators. The results show that the employ of modified gluon propagators is quite valuable in phenomenology, although requires some caution, since the model used to construct the scattering amplitude is a first approximation. The results point out to a gluon propagator with dynamical mass and frozen running coupling constant, although the present results in addition with the previous ones\cite{amn02,amn04,hkn93,gds03} do not definitely determine the most suitable expression for the propagator, mainly due the approximations employed and the still poor knowledge on the interplay of the whole aspects involved in these calculations. %The consistency of the ideas used here can be checked in another type of diffractive processes which should deserve some analysis in this framework. 

%%%%%%%%%%%%%%%%%%%%%%%%%%%%%%%%%%%%%%%%%%%%%%%%%%%%%%%%%%%%%%%%%%%%%%%%%%%%%%%%%%%%%%%%%%%%
%%%%%%%%%%%%%%%%%%%%%%%%%%%%%%%%%%%%%%%%%%%%%%%%%%%%%%%%%%%%%%%%%%%%%%%%%%%%%%%%%%%%%%%%%%%%
\section*{Acknowledgments}
This work was supported partly by the Conselho Nacional de Desenvolvimento Cient\'{\i}fico e Tecnol\'ogico (CNPq). %W.S. thanks the enlightenments provides by referees of this work.

\bibliography{bibdefs,bfkl,difracao,eds,livros,ln,mesvec,outros,pp,qcd,rede,decmff,gg}
\bibliographystyle{unsrt}

\end{document}